# Adoption of TikTok as a Learning Tool in Physical Education: Evidence from the Philippines

Vanessa B. Sibug[1], Jan Henry B. Sunga[1], Emerson Q. Fernando[1]
Roe Vincent S. Ovejas[1], Arjan Gil S. Mendoza[1], Trisha Anne A. Onofre[1]
Agnes R. Regala[1], John Paul P. Miranda[1*]

1. **Pampanga State University**, Pampanga, Philippines

**\* Correspondence:**
John Paul P. Miranda, Pampanga State University, jppmiranda@pampangastateu.edu.ph

## ABSTRACT

This study examines the factors that influence the adoption of TikTok as a learning tool for physical education (PE)-related content among tertiary students in the Philippines. The study applies the Technology Acceptance Model (TAM) and Uses and Gratification Theory (UGT) to assess Information Seeking, Personal Identity, Social Interaction, Entertainment, Perceived Usefulness (PU), Perceived Ease of Use (PEOU), and Intention to Use (IU). A cross-sectional design and Structural Equation Modeling (SEM) were employed. The sample included 1,075 regular TikTok users with an average age of 19 years, the majority of whom were female. The analysis revealed that PU and PEOU were the strongest predictors of IU TikTok for PErelated content. The results indicate that TikTok provides an engaging and accessible medium that supports active learning and participation in PE. The study offers empirical evidence from the Philippines and contributes to the academic discussion on the role of short-form video platforms in PE.

*Keywords: social media, uses and gratification theory, education, TAM, mobile learning, Filipino*

## INTRODUCTION

TikTok is a digital platform that allows users to create and share short videos. It has reached a global user base that exceeds one billion across more than 160 countries (Ceci, 2024). The platform attracts young audiences through brief, creative, and visually oriented content that maintains attention and participation (López-Carril, Watanabe, et al., 2024). Within education, social media applications such as TikTok, YouTube, and Instagram function as instructional tools that support active learning and microlearning (Lamimi et al., 2024; Literat, 2021; Nur Ilianis Adnan et al., 2024). A study on teenagers reported that many users rely on TikTok for school-related content and academic material and that this habit reflects a preference for accessible and concise learning formats (Adelhardt & Eberle, 2024). These developments indicate that social media platforms can hold a role in the delivery of structured content in formal and informal education (Adelhardt & Eberle, 2024; Sari et al., 2022).

Physical education (PE) requires instructional strategies that emphasize observation, demonstration, and performance. Active learning tools are important in this context because they encourage students to apply theoretical knowledge through action and reflection. TikTok provides such opportunities through its visual and participatory format. It enables students to view movement demonstrations, examine technique, and share their own PE-related content (Arranz et al., 2023; Garcia, 2025; Ismail et al., 2021; López-Carril, González-Serrano, et al., 2024; López-Carril, Watanabe, et al., 2024; Nur Ilianis Adnan et al., 2024; Sari et al., 2022). Teachers may incorporate short videos to illustrate proper form, present routines, or provide remote evaluation of student performance (Zhu et al., 2022). These practices correspond with active learning principles in PE that support engagement, autonomy, and collaboration (Garcia, 2025).

International studies have examined TikTok in higher education (Y. Yang et al., 2025) and in sport-related courses (López-Carril, González-Serrano, et al., 2024), but very few studies have focused on its academic integration in the Philippines (Garcia, 2025; Masangcay et al., 2024; Matitu & Santiago, 2023; Tunac et al., 2025). The Philippines presents a distinct context because of strong social media activity, varied technology access, and a youth population that engages intensely with online platforms. Limited attention has been given to TikTok as a learning tool for PE-related content among Filipino tertiary students. PE subjects depend on visual demonstration and repeated practice, and therefore digital platforms that support video-based material may contribute to instruction. The present study addresses this gap by identifying the variables that influence the adoption of TikTok as a learning tool for PE-related content among tertiary students in the Philippines. The study applies the Technology Acceptance Model (TAM) and the Uses and Gratification Theory (UGT) to explain intention to use (IU) TikTok through perceptions of usefulness, ease of use, and motivational factors.

## THEORETICAL FRAMEWORK

TikTok has rapidly emerged as a leading social media platform among Generation Z (Gesmundo et al., 2022), especially for disseminating educational content in PE-related contexts such as dance, exercise, diet, and sports. Research indicates that TikTok can effectively facilitates the sharing and absorption of educational materials which can significantly enhance its perceived usefulness (PU) for educational purposes (Al-maroof et al., 2021; Ismail et al., 2021; Lamimi et al., 2024; Zulkifli et al., 2022). For instance, (2021) suggest that TikTok supports academic success and fosters social-emotional well-being which increases its PU among students. In a separate study, researchers focused on the perspective of teachers regarding the use of TikTok in PE, they concluded that TikTok has a value particularly in promoting active learning and engagement among their students. This supports the notion that information seeking (IS) leads to higher PU of the platform for educational content (Sari et al., 2022). This leads to the following hypothesis: *H1: IS positively influences PU of TikTok for PE-related content*.

The concept of personal identity (PI) is pivotal in shaping user engagement on TikTok. Users frequently customize their profiles and content to reflect their interests and identities which fosters a stronger connection to educational materials that resonate with their self-perception (Fin et al., 2017; Sari et al., 2022). By enabling users to express their identities through content creation, TikTok increases their appreciation and PU of educational videos related to PE (López-Carril, González-Serrano, et al., 2024). This alignment nurtures a sense of belonging and relevance which can enhance the utility of the platform for educational purposes. Thus, this study proposed that: *H2: PI positively influences PU of TikTok for PE-related content.*

The emphasis on TikTok in terms of fostering social interaction (SI) is particularly beneficial in an educational context (Bopp & Stellefson, 2020; Matias et al., 2018; Salasac & Lobo, 2022). The platform encourages communication and relationship-building through shared content which can facilitates collaborative discussions on various educational topics (Bopp & Stellefson, 2020; Salasac & Lobo, 2022). This interactive environment promotes engagement and amplifies the PU of the content and may allow students to benefit from shared learning experiences (Liao, 2021; Salasac & Lobo, 2022). Therefore, the following hypothesis is proposed: *H3: SI positively influences PU of TikTok for PE-related content and H5: SI positively influences Perceived Ease of Use (PEOU) of TikTok for PE-related content*.

Entertainment (ENT) is a fundamental feature of TikTok that captivates users including students in PE contexts (Liao, 2021). Its engaging short-form videos effectively blend ENT with informative content and boost the overall learning experience (Liao, 2021). This entertaining aspect increases user engagement and boosts the PU of the platform which makes it more likely for students to retain and internalize information presented in an enjoyable format (Chaudhary, 2010; Long & Nie, 2021). Research consistently shows that integrating ENT into educational content enhances user interest and retention (Chaudhary, 2010). This positions TikTok as a valuable learning resource for PE. The following hypotheses are proposed:*H4: ENT positively influences PU of TikTok for PE-related content and H6: ENT positively influences PEOU of TikTok for PE-related content*.

Users tend to view a platform as valuable when it is easy to navigate which enhances its PU (Sari et al., 2022). TikTok's intuitive design and user-friendly features enrich users' learning experiences (Liao, 2021). This concept is supported by the TAM which posits that ease of use leads to greater perceived benefits. When students find TikTok easy to engage with, they are more inclined to utilize it regularly for educational purposes (Lamimi et al., 2024; Langreo, 2022). Additionally, PU is a strong predictor of technology adoption (Al-Hattami, 2023; Granić, 2023; Liesa-Orús et al., 2023). It indicates that students are more likely to use TikTok if they believe it provides valuable educational content (Conde-Caballero et al., 2024; Ding et al., 2023; Escamilla-Fajardo et al., 2021). This perceived value reinforces their intention to continue using the platform for learning. Thus, this study proposed the following hypotheses: *H7: PEOU positively influences PU of TikTok for PE-related content, H8: PEOU positively influences the IU TikTok for PE-related content, and H9: PU positively influences the IU TikTok for PE-related content*.

**Figure 1.**
*Proposed hypothesized paths*

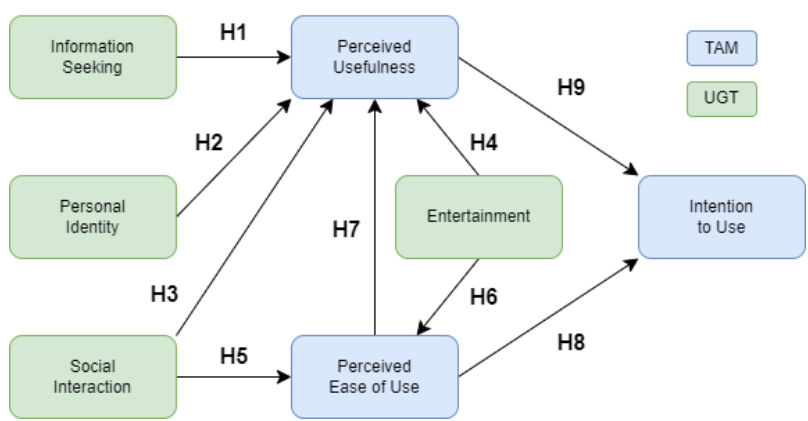

# METHOD

This study employed a cross-sectional design and Structural Equation Modeling (SEM) to examine factors influencing the use of TikTok for PE-related content among tertiary students in Central Luzon, Philippines. The research model was developed based on constructs from the TAM and UGT with relationships hypothesized from relevant literature. A survey instrument measuring seven constructs was utilized: IS, PI, SI, ENT, PU, PEOU, and IU. The survey also gathered demographic data and TikTok usage patterns and a pilot test confirmed the reliability of the instrument with Cronbach's alpha values exceeding 0.70.

Data collection took place from March to May 2024 involving 1,075 valid responses from tertiary students recruited through convenience and snowball sampling. The respondents, primarily female (66.7%, $n$ = 717) with an average age of 19, were regular TikTok users, with 81.4% ($n$ = 875) using the platform daily. Descriptive statistics were calculated using Python programming while SEM analysis was conducted using Python Semopy package to test relationships between the variables. Model fit was assessed using goodness-of-fit indices such as Chi-square, RMSEA, CFI, and TLI to ensure a thorough evaluation of the hypothesized relationships (Doğan, 2022; Hsu et al., 2015).

# RESULTS AND DISCUSSION

For the measurement model results, the construct reliability (CR) for IS (CR = 0.771), PI (CR = 0.758), SI (CR = 0.747), ENT (CR = 0.778), PU (CR = 0.700), PEOU (CR = 0.822), and IU (CR = 0.777) all surpasses the recommended threshold of 0.70, with CR values ranging from 0.700 to 0.822. These results indicated a good internal consistency and ensure the reliability of the measurement scales. The Average Variance Extracted (AVE) for each construct ranges from 0.509 to 0.607. The Average Shared Variance (ASV) ranges from 0.306 to 0.393 and the Maximum Shared Variance (MSV) ranges from 0.393 to 0.469. The measurement model further confirms convergent validity, as the AVE for each construct is greater than both the ASV and MSV (AVE > 0.50; AVE > ASV; AVE > MSV).

Table 1 reports the means and standard deviations (M ± SD) along with the initial and final factor loadings. The results indicated that students show a higher IU TikTok for PE-related content, as reflected in the IU scores (4.7 ± 0.87). In the initial SEM model as illustrated in Figure 2, all constructs reported a significant value of $p$ < 0.05. This indicated that each construct was a significant predictor of the IU TikTok for PE-related content. To improve the model fit, one measurement item from SI and PU were removed in the final model. All in all, there were 19 measurement items in the final model.

Studies have shown that social and educational motivations can facilitate the use of TikTok as an educational tool in PE contexts. For instance, (2021) previously demonstrated that TikTok can impact engagement and educational quality in education. (2021) indicated that an increase in student interest and overall improvement in teaching quality when multimedia communications methods such as TikTok is applied in PE education. This supported the present study where PEOU and PU can be an indicative that TikTok can be a useful tool in PE. Similarly, (2022) found that TikTok can facilitates student engagement in rythmic activities and may be applied in other fields of PE and sports. Moreover, (2022) detailed the trends on how UGT and social media like TikTok can drive user satisfaction and engagement among users. Like many social media platforms, TikTok is being use by many users to find ENT, information, and social interact with other people. This insights from (2022) is parallel with the constructs in this study where IS and SI is seen as a driving force why there is an positive IU TikTok for PE. In addition, study like (2022), shows how short-video platforms is perceived as fun and easy to use which helps in improving user use intention.

**Figure 2.**
*The initial SEM model*

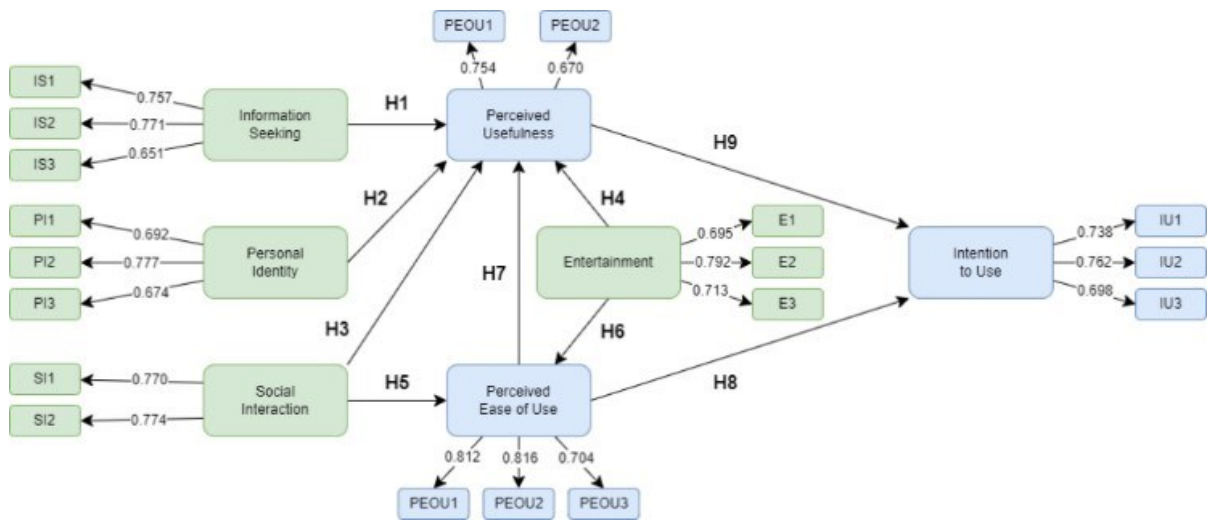

**Table 1.**
*Descriptive statistics and factor loading*

| Construct | Variables | M ± SD | Factor Loadings | |
|---|---|---|---|---|
| | | | Initial | Final |
| IS | IS1 | 4.82 ± 0.83 | 0.751 | 0.757 |
| | IS2 | 4.72 ± 0.91 | 0.768 | 0.771 |
| | IS3 | 4.47 ± 1.02 | 0.647 | 0.651 |
| PI | PI1 | 4.68 ± 0.88 | 0.687 | 0.692 |
| | PI2 | 4.71 ± 0.82 | 0.768 | 0.777 |
| | PI3 | 4.75 ± 0.86 | 0.669 | 0.674 |
| SI | SI1 | 4.61 ± 0.96 | 0.758 | 0.770 |
| | SI2 | 4.22 ± 1.19 | 0.786 | 0.774 |
| | SI3 | 4.65 ± 0.85 | 0.467 | - |
| ENT | ENT1 | 4.87 ± 0.76 | 0.693 | 0.695 |
| | ENT2 | 4.84 ± 0.83 | 0.784 | 0.792 |
| | ENT3 | 4.77 ± 0.81 | 0.707 | 0.713 |
| PU | PU1 | 4.71 ± 0.81 | 0.723 | 0.754 |
| | PU2 | 4.75 ± 0.80 | 0.702 | 0.670 |
| | PU3 | 4.73 ± 0.80 | 0.564 | - |
| PEOU | PEOU1 | 4.83 ± 0.80 | 0.81 | 0.812 |
| | PEOU2 | 4.82 ± 0.80 | 0.812 | 0.816 |
| | PEOU3 | 4.82 ± 0.76 | 0.695 | 0.704 |
| IU | IU1 | 4.64 ± 0.92 | 0.734 | 0.738 |
| | IU2 | 4.76 ± 0.85 | 0.748 | 0.762 |
| | IU3 | 4.70 ± 0.84 | 0.692 | 0.698 |

In Table 2, the inter-construct correlations along with the square root of AVE is presented. They all ranged from 0.401 to 0.685. Since a few constructs exceeded the 0.60 threshold, the Variance Inflation Factor (VIF) was measured to assess the severity of multicollinearity among the constructs. The results showed that the VIF values ranged from 1.99 to 2.68 and the tolerance values ranged from 0.37 to 0.50. These results indicated that multicollinearity was not a concern in the dataset. Furthermore, when goodness-of-fit indices were taken into account, it is observed in this study that the model achieved a satisfactory fit across key indicators, including RMSEA,

**Sibug et al. (2026)**

NFI, CFI, GFI, TLI, and AGFI. Specifically, the Root Mean Square Error of Approximation (RMSEA) of 0.041, which is below the recommended threshold of 0.05, indicates a good fit with minimal approximation error. This confirms the robustness of the model. Similarly, the Normed Fit Index (NFI) of 0.970 falls within the acceptable range for a good fit model which indicates a significant improvement over the null model and assumes no relationships among the variables. The Comparative Fit Index (CFI) of 0.981 further supports the adequacy of the model and demonstrates its effectiveness in explaining a large proportion of the variance in the observed data compared to a baseline model. The results are consistent with the study of (2015) where they specified that RMSEA, CFI, and TLI are good measurements for detecting within-group misspecifications (i.e., having minimal approximation error and high model fit).

Additional support for the model fit comes from the Goodness-of-Fit Index (GFI) value of 0.970. This indicates that the model accounts for a significant portion of the variance-covariance matrix. Furthermore, the Tucker-Lewis Index (TLI) of 0.976 indicates the comparative strength of the model in which the any values closer to 1.00 indicates a good fit against the baseline. Furthermore, the Adjusted Goodness-of-Fit Index (AGFI), which adjusts for model complexity, stands at 0.963. This reinforce the suitability of the model after considering the number of parameters estimated. Collectively, these indices fall within recommended thresholds for a well-fitting model, signifying strong explanatory power and a reliable representation of the underlying data structure (Doğan, 2022; Hsu et al., 2015).

**Table 2.**
*Inter-construct correlations with square root of AVE*

|  | **IS** | **PI** | **SI** | **E** | **IU** | **PEOU** | **PU** |
|---|---|---|---|---|---|---|---|
| **IS** | 0.728 | | | | | | |
| **PI** | .685 | 0.716 | | | | | |
| **SI** | .627 | .623 | 0.772 | | | | |
| **ENT** | .624 | .641 | .577 | 0.734 | | | |
| **IU** | .609 | .636 | .545 | .626 | 0.733 | | |
| **PEOU** | .498 | .530 | .401 | .577 | .675 | 0.779 | |
| **PU** | .543 | .610 | .515 | .630 | .662 | .606 | 0.713 |

Table 3 summarizes the hypotheses testing, with six out of nine hypothesized paths supported. The $R^2$ values indicates the proportion of variance in the dependent variables explained by the independent variables. A higher value for $R^2$ signifies a better model fit. Results showed that the model explained 62.66% of the variance in IU, 55.74% in PU, 52.20% in PEOU, 57.54% in ENT, 49.80% in SI, 61% in PI, and 57.85% in IS. The over 50% variances indicates the robustness in capturing key factors influencing TikTok adoption for PE-related content (Gumilar et al., 2019). The strongest relationship observed is between PU and IU (H9), with a path coefficient of 0.966. It also shows that PU is a key driver of IU. This is consistent with the foundational insights of the TAM (Davis, 1989; Hua & Chiu, 2022; Tennakoon et al., 2023). In addition, ENT has a significant impact on PEOU (H6) with a coefficient of 0.640. This indicated that ENT strongly influences PEOU. Additionally, PEOU plays a crucial role in shaping both PU and IU and emphasized its importance in the intent to use process. For instance, this result mirror the findings by (2020) where they found that PEOU impacts both PU and IU among users. While SI significantly affects PU, it does not have significant influence on PEOU. This suggests that social factors impact perceptions of usefulness but not ease of use. This pattern suggests that while social factors may alter users' perceptions, they do not necessarily change how the users find the platform easy to use. This distinction also suggest that social endorsement can increase the perceived benefits of TikTok but does not necessarily make the platform appear simpler or easier to use. Figure 3

**Sibug et al. (2026)**

illustrates the final model for evaluating the factors affecting the IU TikTok for PE-related content among tertiary students.

**Table 3.**
*Hypothesis testing results*

| H# | Structural Paths | Standardized Path Coefficients | p-value | Result |
|---|---|---|---|---|
| H1 | IS ➔ PU (+) | 0.025 | 0.64 | Not supported |
| H2 | PI ➔ PU (+) | 0.237 | 0.00 | Supported |
| H3 | SI ➔ PU (+) | 0.10 | 0.02 | Supported |
| H4 | ENT ➔ PU (+) | 0.064 | 0.064 | Not supported |
| H5 | SI ➔ PEOU (+) | 0.064 | 0.14 | Not supported |
| H6 | ENT ➔ PEOU (+) | 0.640 | 0.00 | Supported |
| H7 | PEOU ➔ PU (+) | 0.325 | 0.00 | Supported |
| H8 | PEOU ➔ IU (+) | 0.217 | 0.00 | Supported |
| H9 | PU ➔ IU (+) | 0.966 | 0.00 | Supported |

The unsupported hypotheses (H1, H4, and H5) demonstrate that not every motivational or social factor exerts a direct influence on adoption of TikTok for PE-related content. IS (H1) did not show a significant effect on PU, which implies that students already consider TikTok a familiar and accessible medium rather than a primary information source. ENT (H4) did not significantly affect PU, which indicates that enjoyment alone does not determine the platform's academic value. SI (H5) did not significantly affect PEOU, suggesting that ease of operation is individually assessed rather than socially reinforced.

**Figure 3.**
*The final SEM model*

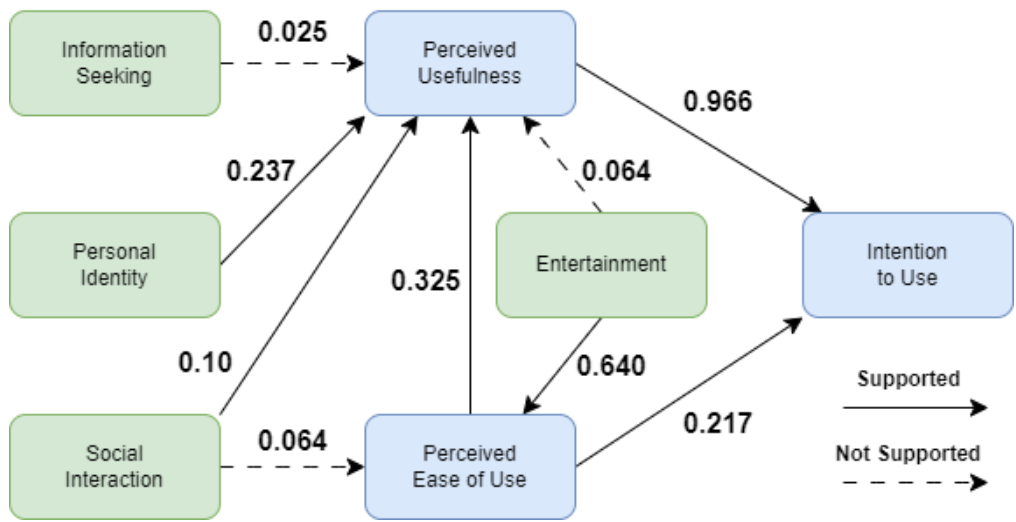

## CONCLUSION AND FUTURE WORKS

primary variables that influenced IU. SI contributed to the perception of usefulness, while accessibility and enjoyable content supported ease of use. These relationships indicate that TikTok can function as an effective medium for promoting PE-related content and increasing student participation in digital instruction. The integration of social media within PE instruction may enhance motivation, enrich classroom experience, and align with the objectives of contemporary education.

**Sibug et al. (2026)**

The integration of TikTok into formal instruction also requires ethical and pedagogical consideration. Ethical responsibility involves the protection of student data and the assurance that uploaded materials comply with content standards and institutional guidelines. Teachers and administrators must establish clear policies for the responsible use of social media to prevent privacy issues and misuse. Pedagogically, TikTok should be applied in ways that strengthen learning outcomes, maintain discipline relevance, and support accurate demonstration of physical skills. Activities should remain structured and directly related to curriculum objectives to sustain educational value.

Further investigation that includes the perspectives of teachers and administrators could extend understanding of how social media platforms influence instructional practice in PE. Future studies may also employ frameworks such as UTAUT2 or other contemporary adoption models to analyze similar contexts and examine longer-term effects on learning engagement and performance.